\documentclass[10pt]{article}

\usepackage{amsmath}
\usepackage{amssymb}
\usepackage{graphicx}
\usepackage{cite}
\usepackage{color} 
\usepackage{caption}
\usepackage{subcaption}

\makeatletter
\renewcommand{\@biblabel}[1]{\quad#1.}
\makeatother

\date{}

\def\mean#1{\langle#1\rangle}

\begin{document}

\begin{flushleft}
{\Large
  \textbf{Effect of voltage dynamics on response properties in a model of sensory hair cell} }
\\
\bigskip
Rami Amro$^1$ and  Alexander B. Neiman$^2$\\
\bigskip
{\em Department of Physics and Astronomy, Ohio University, Athens OH 45701, USA}\\
$^1$ ramiamro@hotmail.com\\
$^2$ neimana@ohio.edu\\
\end{flushleft}

\section*{Abstract}
Sensory hair cells in auditory and vestibular organs rely on active mechanisms to achieve high sensitivity and frequency selectivity. Recent experimental studies have documented self-sustained oscillations in hair cells of lower vertebrates on two distinct levels. First, the hair bundle can undergo spontaneous mechanical oscillations. Second, somatic electric voltage oscillations across the baso-lateral membrane of the hair cell have been observed.  We develop a biophysical model of the bullfrog's saccular hair cell consisting of  two compartments, mechanical and electrical,  to study how the  mechanical and the voltage oscillations interact to produce coherent self-sustained oscillations and how this interaction contributes to the overall sensitivity and selectivity of the hair cell. The model incorporates nonlinear mechanical stochastic hair bundle system coupled bi-directionally to a Hodgkin-Huxley type system describing somatic ionic currents.  We isolate regions of coherent spontaneous oscillations in the parameter space of the model and then study how coupling between compartments affect sensitivity of the hair cell to external mechanical perturbations. We show that spontaneous electrical oscillations may enhance significantly the sensitivity and selectivity of the hair cell.

\section*{Introduction}

Hair cells are peripheral sensory receptors in the inner ear of vertebrates which transduce mechanical vibrations to electrical signals \cite{Fettiplace2006a,Hudspeth2008}. External mechanical perturbations are translated to electrical signals through the  
mechano-electrical transduction (MET) channels located in the  stereocilia, apical projections of the cell immersed in potassium (K) rich endolymph. Stereocilia are arranged in the hair bundle in rows of increasing heights and linked by the so-called tip links. 
External mechanical perturbations deflect the hair bundle resulting in changing of the tension in the tip links, which may elicit opening or closing of MET channels, mediating the influx of potassium and calcium ions to the cell and thus changing the cell's potential. Experimentally observed extreme sensitivity and frequency selectivity of hair cells suggested several mechanisms of active amplifications  \cite{Fettiplace2006a,Hudspeth2008,Ashmore2010}. One of these mechanisms relies on active dynamics of the hair bundle  which can be placed at the verge of dynamical instability, i.e. Andronov-Hopf bifurcation, providing the cell with high sensitivity, sharp frequency tuning curve and a compressive nonlinearity \cite{Eguiluz2000,MarHud99,Hudspeth2010}. MET channels in stereocilia opening in concert exert forces on the hair bundle, the phenomenon known as gating compliance \cite{Howard1988}, resulting in a nonlinear force-displacement relation. In some hair cells the gating compliance may result in a differential negative stiffness of the hair bundle leading to its instability \cite{Martin2000}. Combined with myosin-mediated adaptation \cite{Eatock2000,Fettiplace2003} this nonlinearity result in spontaneous mechanical hair bundle oscillatory motion. In particular,  
spontaneous oscillations of the hair bundle  have been observed in many lower vertebrates such as amphibian sacculus, or turtle basilar paplia \cite{Manley1988a,Manley1997,Manley2001,Martin2003,Ramunno-Johnson2009} and presumably responsible for the phenomenon of otoacoustic emissions in nonmammals \cite{Manley1997,Manley1988a,Gelfand2010}. Importantly, spontaneous hair bundle oscillations are essentially stochastic due to several thermal noise sources \cite{Nadrowski2004}. This natural stochasticity can be reduced in coupled hair bundles \cite{Dierkes2008,Barral2010}.

Intriguingly, the same type of hair cells  show spontaneous oscillations of the membrane potential. For example, several experimental groups reported on spontaneous voltage oscillations in saccular hair cells \cite{Ospeck2001,Catacuzzeno2004,Jorgensen2005,Rutherford2009}. The functional role of these large-amplitude voltage oscillations is currently not known. One suggestion is that voltage oscillations provide additional feedback to the hair bundle dynamics which may help to reduce thermal fluctuations and thus to improve sensitivity of the hair cell  \cite{Han2010}. Indeed earlier studies reported on reverse electro-mechanical transduction, providing evidence for bi-directional coupling between the hair bundle complex and the potential of the cell body \cite{Denk1992}. Furthermore, transepithelial electrical stimulation of hair cells elicits the hair bundle motion \cite{Bozovic2003}. Finally, a recent experimental study showed that somatic ionic conductances affect significantly the mechanical oscillations of hair bundles. \cite{Ramunno-Johnson2010}.

In this paper we develop a model of sensory hair cell which includes mechanical and electrical compartments coupled bidirectionally.
Sensitivity, frequency selectivity and compressive nonlinearity are three main  features of the hair bundle dynamics \cite{Hudspeth2010}.
We use the model to study how the coupling affects these three feature of the hair cell dynamics.

\section*{Two compartmental model of a hair cell}

Our two compartmental system is based on previously developed models for the hair bundle mechanics and for the dynamics of the cell potential of the bullfrog's saccular hair cells. The coupling between the two compartments is bidirectional. The coupling from the mechanical to electrical compartment is due to 
direct mechano-electrical transduction, i.e. due to the MET current \cite{Holton1986,Pickles1992}. The coupling from the electrical to mechanical compartment is mediated by the reverse electro-mechanical transduction whereby the cell voltage affects the calcium ions concentration near the myosin molecular motors controlling the rate of adaptation of the hair bundle.  
  
Several simple models were proposed to account for the active hair bundle motion including spontaneous mechanical  oscillations  in the bullforg's sacculus \cite{Choe1998,Martin2003,Nadrowski2004,Tinevez2007,Roongthumskul2011,OMaoileidigh2012}. In this paper we use a model proposed in \cite{Nadrowski2004,Tinevez2007}. The overdamped dynamics of the hair bundle is described by two stochastic differential equations: first  for the position of the bundle tip, $X$ and   second  for the displacement of the myosin motors, $X_\mathrm{a}$ along  the stereocilia:
\begin{eqnarray}
\label{mech.eq}
&&\lambda \frac{dX}{dt}=-K_\mathrm{GS}(X-X_\mathrm{a}-D\,P_\mathrm{o})-K_\mathrm{SP}X+F_\mathrm{ext}(t)+\sqrt{2k_\mathrm{B}T\lambda}\eta(t),  \\
&&\lambda_\mathrm{a} \frac{dX_\mathrm{a}}{dt}=K_\mathrm{GS}(X-X_\mathrm{a}-D\,P_\mathrm{o})-F_\mathrm{max}(1-S\,P_\mathrm{o})
+\sqrt{3k_\mathrm{B}T\lambda_\mathrm{a}}\eta_\mathrm{a}(t), \nonumber \\
&&P_\mathrm{o}(X,X_\mathrm{a})=\frac{1}{1+A\,e^{-(X-X_\mathrm{a}) K_\mathrm{GS} D/(N k_\mathrm{B} T)}},\quad
A=e^{[\Delta G +K_\mathrm{GS}D^2/(2N)]/(k_\mathrm{B} T)}. \nonumber
\end{eqnarray}
In Eqs.(\ref{mech.eq}) $\eta(t)$ and $\eta_\mathrm{a}(t)$ are uncorrelated Gaussian white noise sources modeling thermal noise due to Brownian motion, MET channel cluttering and stochastic binding and unbinding of adaptation motors \cite{Nadrowski2004}. The external stimulus enters the model via the external force term $F_\mathrm{ext}(t)$.
Definition of parameters and their values are given in Table~\ref{rasymb} and are essentially the same as used in \cite{Dierkes2008,Nadrowski2004}. 
\begin{table}[hbtp]
 \caption{Parameters of the mechanical compartment}
\label{rasymb} 
\centering                         
\begin{tabular}{|l |l |l |l|} 
\hline       
Symbol & Definition & value  \\ [0.5ex] 
\hline 
$\lambda$ & Drag coefficient of the hair bundle & 2.8 $\mu$Ns/m  \\ 
$\lambda_\mathrm{a}$ & Effective friction coefficient of the molecular motors & 10.0 $\mu$Ns/m  \\ 
$K_\mathrm{GS}$&  Gating spring stiffness & 0.75 mN/m  \\
$K_\mathrm{SP}$ & Stereocilia pivots stiffness & 0.6 mN/m \\
D & Gating swing of MET channel & 60.9 nm \\
$T$& Ambient Temperature & 300K \\ 
$F_\mathrm{{max}}$& Maximal motors force & 55 pN  \\
$S$ & Dimensionless Ca feedback strength & 1.13  \\
$P_\mathrm{o}$ &   Open probability of MET channels &  \\
$N$ &   Number of MET channels & 50 \\
$\Delta G$ &   Energy change on MET channel opening  & $10 k_\mathrm{B} T$ \\
\hline 
\end{tabular}
\end{table} 
The control parameters of the hair bundle model are $S$ and $F_\mathrm{max}$ which determine the strength of calcium (Ca) influence on adaptation motors and maximal force generated by the motors at stall when all MET channels are closed, respectively.  

The dynamics of the cell's potential can be described by a Hodgkin-Huxley type system developed in \cite{Catacuzzeno2004,Jorgensen2005,Rutherford2009}. We used a modified version of this model described in details in \cite{Neiman2011c}. The model includes 6 basolateral ionic currents and the MET current. The current balance equation is
\begin{equation}
\label{v.eq1}
C_\mathrm{m} \frac{dV}{dt} = -
I_\mathrm{K1}-I_\mathrm{BKS}-I_\mathrm{BKT}-I_\mathrm{DRK}-I_\mathrm{h}-I_\mathrm{Ca}-I_\mathrm{MET},
\end{equation}
where $I_\mathrm{K1}$ is the inwardly rectifier potassium (K) current, $I_\mathrm{BKS}$ and $I_\mathrm{BKT}$ are Ca-activated steady and transient K currents, $I_\mathrm{h}$ is sodium / potassium h-type current, $I_\mathrm{DRK}$ is the direct rectifier K current and $I_\mathrm{Ca}$ is the Ca current. The equation for the potential (\ref{v.eq1}) is accompanied by equations for kinetics of ionic currents listed and for  intracellular 
$[\mathrm{Ca}^{2+}]$, totaling 12 differential equations. 
The detailed description of this system and parameters is provided in \cite{Neiman2011c}. The control parameters of the electrical compartment are the maximal conductance of inwardly rectifier current, $g_\mathrm{K1}$, and relative strength of Ca-activated currents, $b_\mathrm{K}$. Depending on these two parameters the electrical compartment shows rich variety of oscillatory patterns including quasiperiodic, bursting and chaotic oscillations \cite{Neiman2011c}. By setting the fast activation variable for the inwardly rectifier current ($I_\mathrm{{K1}}$) and  $[\mathrm{Ca}^{2+}]$ to their steady states we reduced the dimension of the electrical compartment by 2. We verified that this reduction did not change the voltage dynamics quantitatively.

Forward coupling from the mechanical to the electrical compartment is due to the MET current entering in Eq.(\ref{v.eq1}), 
\begin{equation}
I_\mathrm{MET}=g_\mathrm{MET} P_\mathrm{o}(X,X_\mathrm{a}) (V(t)-E_\mathrm{MET}), 
\label{ramet}
\end{equation} 
where $E_\mathrm{{MET}}=0$ mV is the reversal potential potential and $g_\mathrm{MET}$ is the maximal value of the MET conductance, i.e. when all MET channels are open. The MET conductance $g_\mathrm{MET}$ thus serves as a ``forward`` coupling parameter.  The dynamics of the hair bundle depends crucially on the concentration of Ca ions at the adaptation motors cite  $[\mathrm{Ca}^{2+}]_\mathrm{M}$,  which is determined by the open probability of MET channels and by the potential of the cell $V$ via  electrodiffusion \cite{Martin2003,Tinevez2007}. For example, a decrease of the receptor potential $V$ enhances the influx of Ca ions through MET channels resulting in increase of $[\mathrm{Ca}^{2+}]_\mathrm{M}$ which in turn inhibits the motor activity. This backward 'voltage to mechanics'' coupling can be introduced via voltage-dependent calcium feedback parameter $S(V)$ in Eqs.(\ref{mech.eq}) 
\cite{Han2010},
\begin{equation}
S(V)=S_0\frac{f(V)}{f(V_0)}, \quad f(V)=\frac{\epsilon V}{1-e^{\epsilon V}}, \quad \epsilon=2q_e /(k_\mathrm{B} T), \nonumber
\end{equation}
where $S_0$ is a value of the feedback parameter $S$ at a reference potential $V=V_0=-55$~mV and $q_e$ is the elementary charge. Since the voltage deviations are in the range -80 -- -30 mV, $S(V)$ can is approximated by its first order Taylor expansion around the reference potential,
\begin{equation}
   S(V)=S_0\left[1+\alpha (V/V_0-1)\right], 
\label{S}   
\end{equation}
where we introduced the dimensionless parameter $\alpha$ controlling the strength of the backward coupling, such that $\alpha=0$ corresponds to the hair bundle uncoupled from the variations of the receptor potential.

Two compartmental stochastic model composed of the hair bundle model (\ref{mech.eq}) and of the Hodgkin-Huxley type model for the membrane potential (\ref{v.eq1}) coupled via Eqs.(\ref{ramet}) and (\ref{S}) was studied numerically using the Euler-Maruyama method with the time step 0.1~ms. 
The control parameters of both compartments were chosen such that in the absence of coupling ($\alpha=0$, $g_\mathrm{MET}=0$), stimulus and noise, the hair bundle and the voltage were at a stable equilibrium. In particular we fixed the parameters of the mechanical compartment at $(F_\mathrm{{max}},S)=(55 \:\mathrm{pN},1.13)$, and the parameters of the electrical compartment at $(b_\mathrm{{K}},g_\mathrm{K1})=(0.01 ,1.0\: \mathrm{nS})$. The coupling strengths $g_\mathrm{MET}$ and $\alpha$ were then used as control parameters. Bifurcation analysis of equilibrium states of the deterministic model, i.e. stochastic terms $\eta$, $\eta_\mathrm{a}=0$  in Eqs.\ref{mech.eq}, was performed using CONTENT continuation software package \cite{content}. We then explored spontaneous stochastic dynamics of the model versus the coupling strengths. In particular we calculated the power spectrald density (PSD) of spontaneous hair bundle motion and then measured the quality factor $Q$ of the main peak in the specrum and the mean amplitude of spontaneous oscillations as a function of the coupling strengths $g_\mathrm{MET}$ and $\alpha$.

The sensitivity and frequency selectivity of the hair cell was estimated using periodic external force,
\begin{equation}
F_\mathrm{ext}(t)= F_0\cos(2\pi f_\mathrm{s}t),
\label{Fext}
\end{equation}    
applied to the mechanical compartment (\ref{mech.eq}). We calculated the time dependent averages of the hair bundle position, $\mean{X(t)}$ and of the membrane potential $\mean{V(t)}$ by averaging over 300 realizations of stochastic terms $\eta$ and $\eta_a$ during 500 periods of the external sinusoidal force. We then estimated the mechanican and electrical sensitivities as the ratios of the first Fourier harmonic of these averages,  $\tilde {X}(f_\mathrm{s})$; $\tilde {V}(f_\mathrm{s})$, to the amplitude of the stimulus, $F_0$, 
\begin{equation}
\chi_{f}(f_\mathrm{s}, F_0) = | \tilde {X}(f_\mathrm{s})|/F_0, \quad 
\chi_{V}(f_\mathrm{s}, F_0) = | \tilde {V}(f_\mathrm{s})|/F_0,
\label{chif}
\end{equation} 
in units of nm/pN for the mechanical sensitivity $\chi_{f}$, and mV/pN for the electrical sensitivity, $\chi_{V}$.

\section*{Results and discussion}

\begin{figure}[hbtp]
\begin{center}
\centering
\begin{subfigure}[b]{0.45\textwidth}
\includegraphics[width=\textwidth]{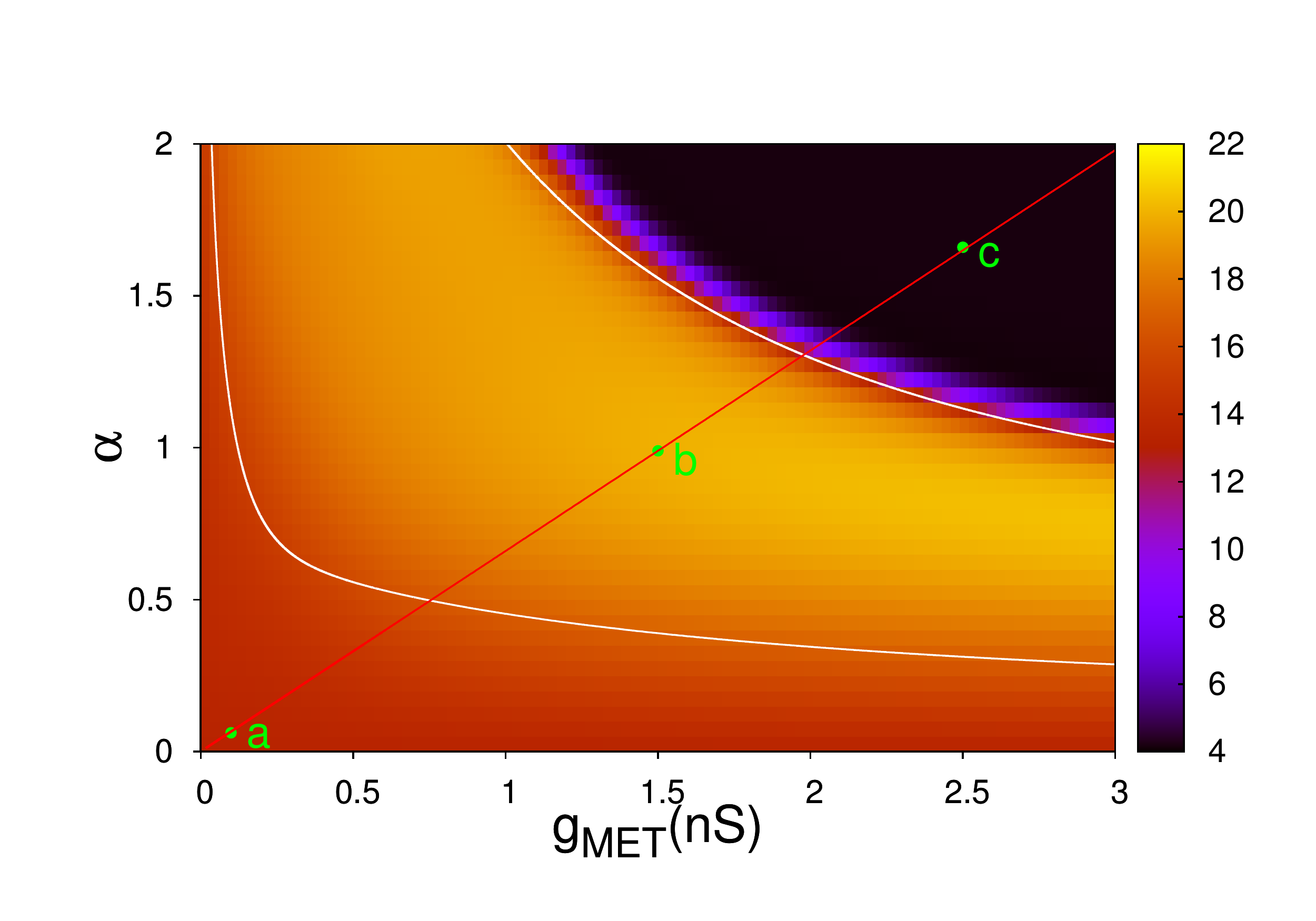}
\end{subfigure}
\begin{subfigure}[b]{0.45\textwidth}
\includegraphics[width=\textwidth]{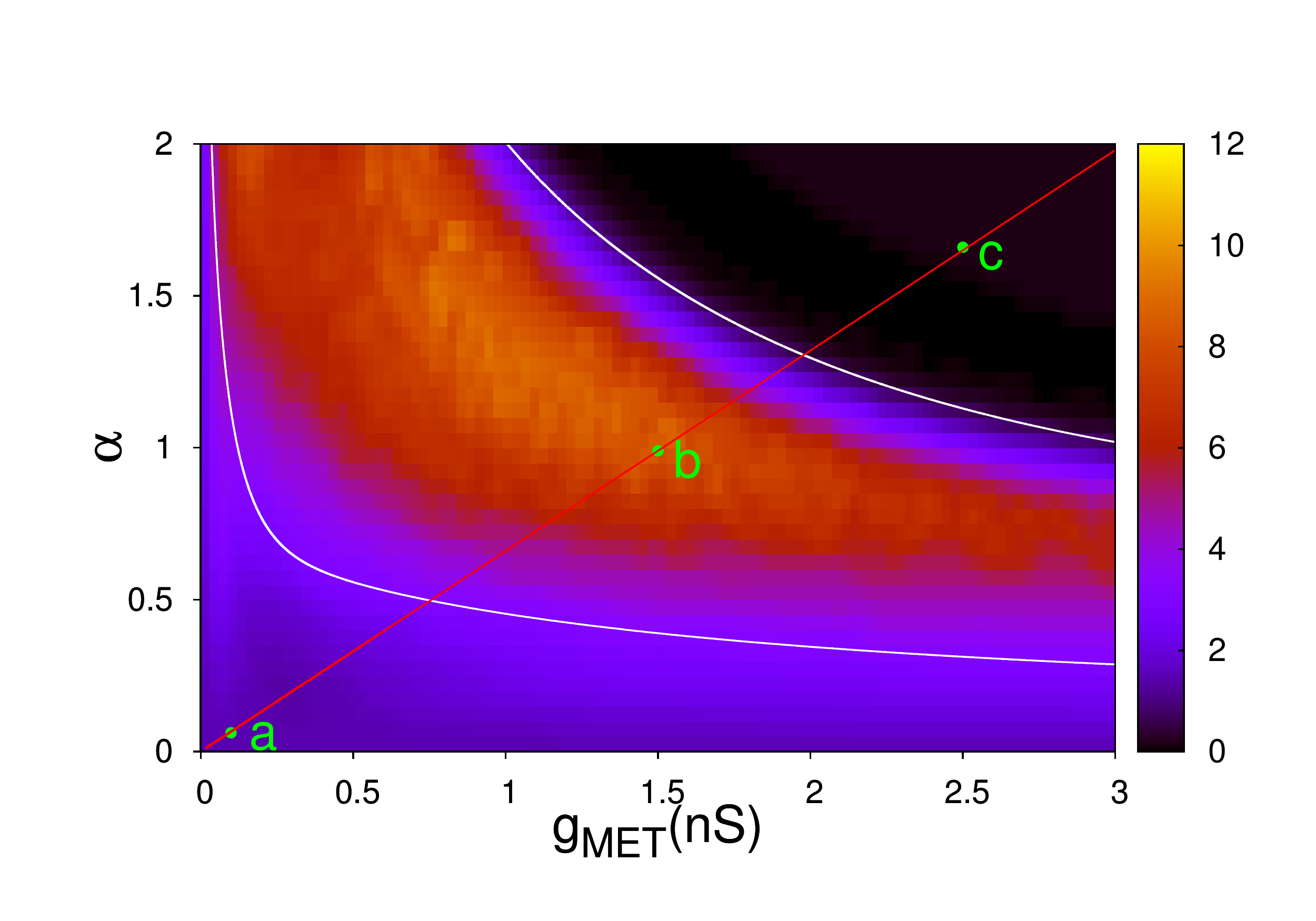}
\end{subfigure}
\end{center}
\caption{Spontaneous dynamics of the hair cell model versus coupling strengths between mechanical and electrical compartments.
Left: Standard deviation (SD) of the hair bundle position, the color bar shows SD in nm.
Right: Quality factor $Q$ of the main peak in the power spectral density of the hair bundle oscillation.
On both panels the AH bifurcation lines of the deterministic system are shown by white lines. Points a, b and c indicate three different regions discussed in the text.}
\label{xsmaps}
\end{figure}

In absence of coupling and noise the deterministic model is at a stable equilibrium, i.e. the mechanical compartment is at equilibrium with most of MET channels closed,  $P_\mathrm{o} \approx 0.19$ and the electrical compartment is at $V=-53.5$~mV. With the increase of coupling the equilibrium bifurcates to a limit cycle via the Andronov-Hopf (AH) bifurcation. On the parameter plane ($g_\mathrm{MET}, \alpha$) the lines of the AH bifurcations isolate a region of spontaneous oscillations, shown in Fig.~\ref{xsmaps}. In the region below the lower AH bifurcation line, most of MET channels are closed, while in the region above the upper AH bifurcation line most of MET channels are open and the cell is depolarized. In the region between the AH bifurcation lines the deterministic model exhibit large-amplitude synchronous oscillations in mechanical and electrical compartments.  

\noindent
Thermal noise induces stochastic hair bundle oscillations even outside the deterministic oscillatory region. Figure \ref{xsmaps} (left panel) shows  that a region of large amplitude mechanical oscillations extends below the lower boundary of deterministic oscillatory region. Noise-induced bundle motion leads to opening of MET channels and since the Ca feedback is relatively weak 
below the lower AH bifurcation line, adaptation brings the bundle back to the equilibrium position allowing  large-amplitude oscillation. 
\begin{figure}[h!]
\begin{center}\includegraphics[scale=0.45]{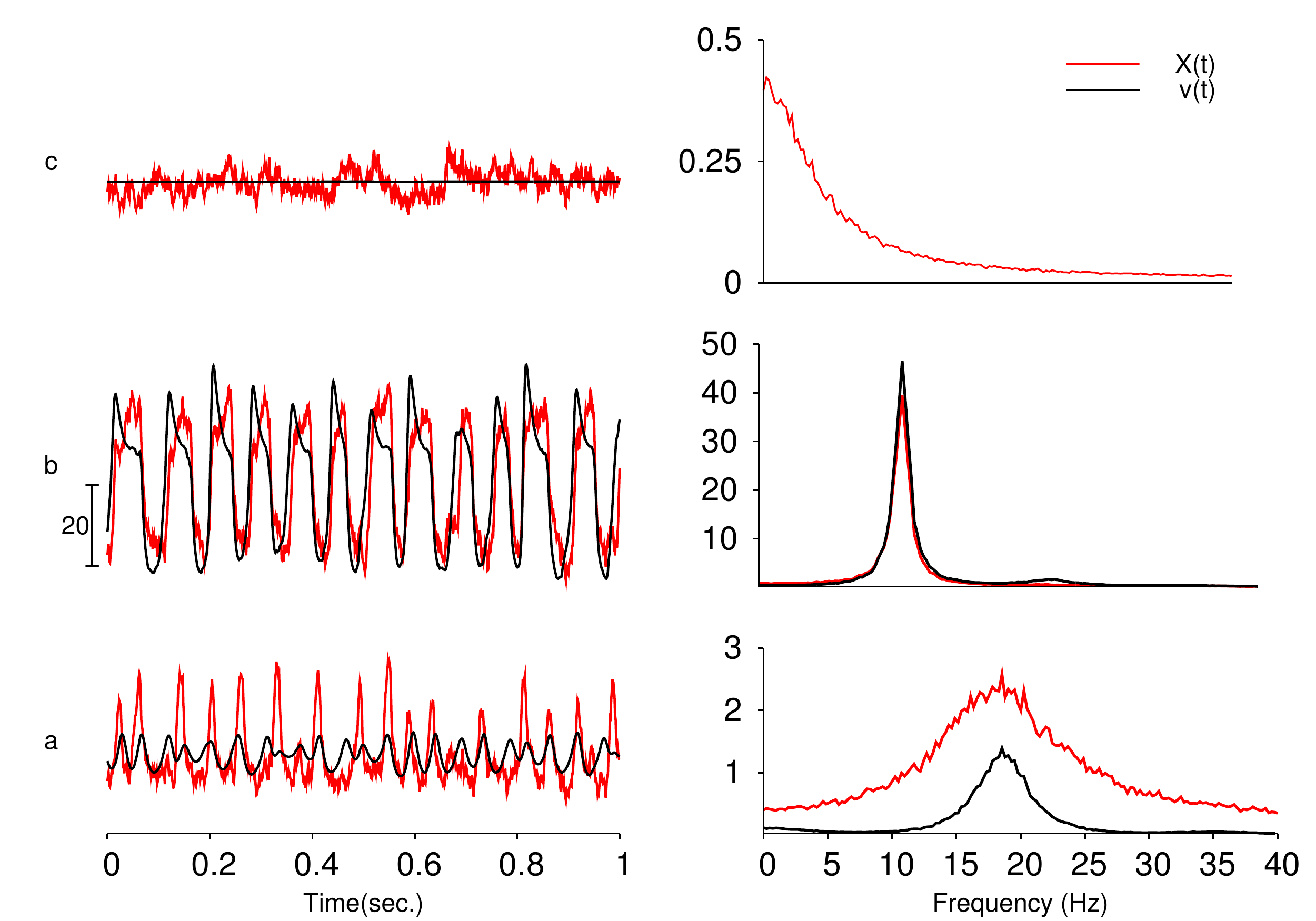}\end{center}
\caption{Time series (left ) and PSDs (right) of  spontaneous mechanical oscillations(red,[nm]), and  membrane potential  (black, [mV]) for the points shown in Fig.~\ref{xsmaps}. Units of PSDs are [nm$^2$/Hz] for the hair bundle displacement and [mV$^2$/Hz] for the membrane potential.}
\label{timeseries}
\end{figure}
On the contrary, above the upper boundary of deterministic oscillatory region large values of the backward coupling $\alpha$ lead to weak adaptation resulting in small-amplitude fluctuations of the hair bundle around the equilibrium with open MET channels. This is illustrated in Fig.~\ref{timeseries} which shows the hair bundle displacement and the voltage traces at three points corresponding to three distinct regions on the parameter plane of the system (cf Fig.~\ref{xsmaps}).

\begin{figure}[hbtp]
\begin{center}
\includegraphics[scale=0.55]{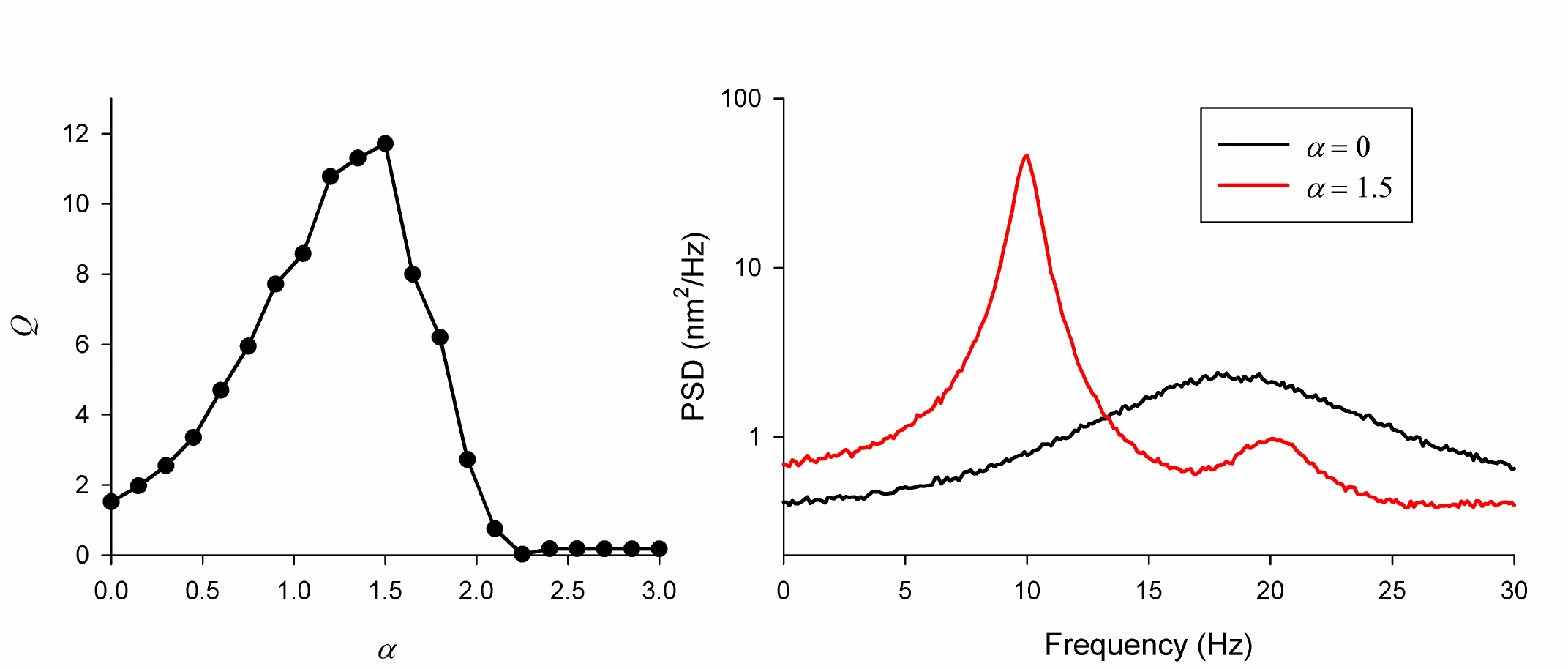}
\end{center}
\caption{Influence of the electrical compartment on the hair bundle oscillation. Left: Quality factor of the mechanical oscillation versus the backward coupling strength $\alpha$ for the fixed value of the MET conductance, $g_\mathrm{MET}=1$~nS. Right: PSD of the hair bundle displacement for indicated values of $\alpha$.} 
\label{op}
\end{figure}

\bigskip
\noindent
The coherence of spontaneous oscillations quantified with the quality factor, $Q$, of the main peak in the PSD of the hair bundle position, shows non-monotonous behavior being maximal at the center of the oscillatory region (point b in the right panel of Fig.~\ref{xsmaps}). In the region below the lower AH line (point a in Figs.~\ref{xsmaps} and \ref{timeseries}) the dynamics of the system is dominated by noise resulting in noisy oscillations around the equilibrium with broad peaks in PSDs at the natural frequency determined by the imaginary parts of the equilibrium's eigenvalues. In the deterministic oscillatory region (point b in 
Figs.~\ref{xsmaps} and \ref{timeseries}) both compartments show synchronous and coherent oscillations with sharp peaks in PSDs. 
Finally, the region above the upper AH line (point c in Figs.~\ref{xsmaps} and \ref{timeseries}) is characterized by overdamped small amplitude fluctuations with a low-frequency Lorentzian type PSD. The effect of the backward electro-mechanical coupling on the hair bundle dynamics is further illustrated in Fig.~\ref{op} showing that the coherence of spontaneous hair bundle oscillations is maximized by the backward coupling $\alpha$. 

\begin{figure}[hbtp]
\centering
\begin{subfigure}[b]{0.45\textwidth}
\includegraphics[width=\textwidth]{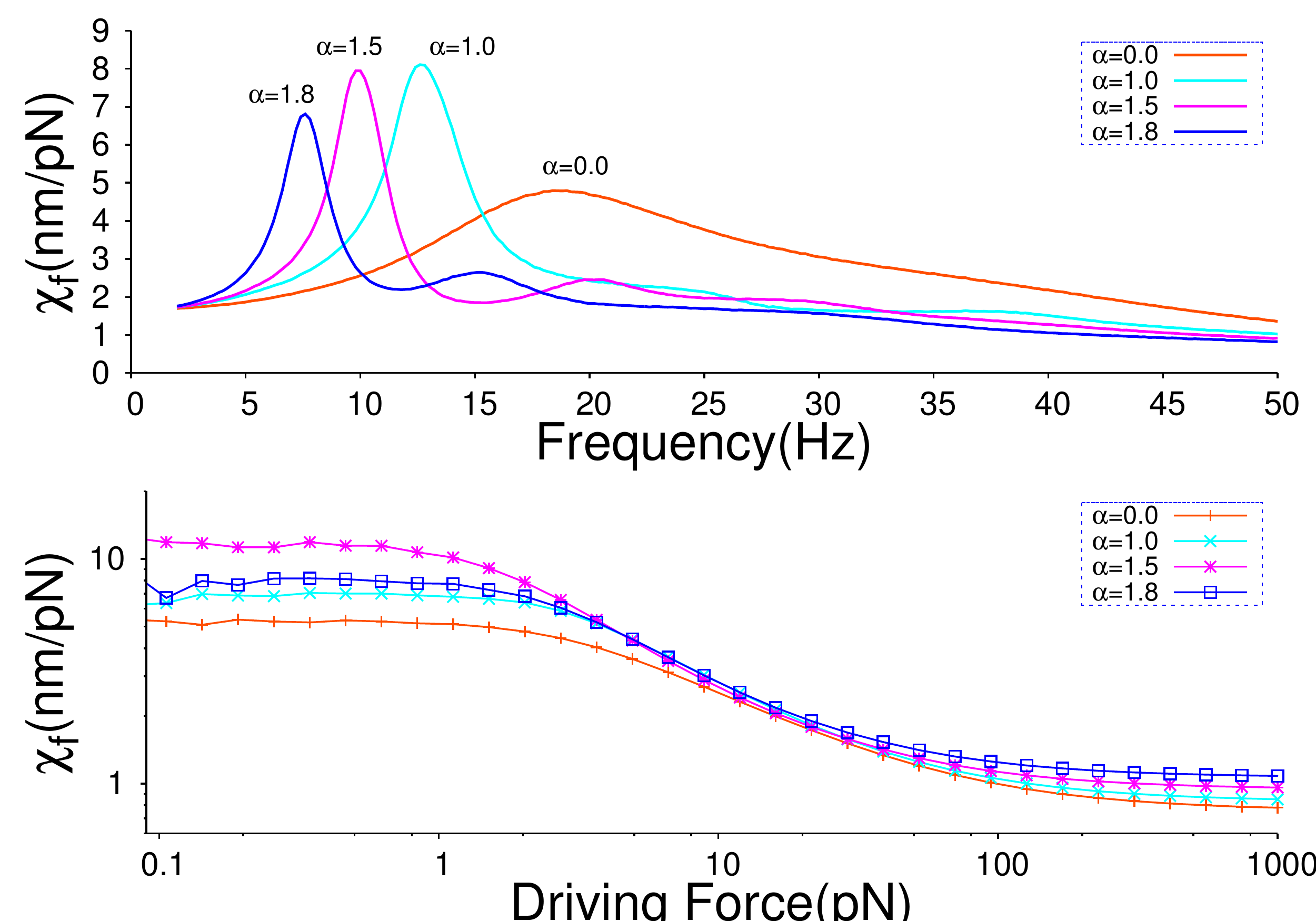}
\end{subfigure}
\begin{subfigure}[b]{0.45\textwidth}
\includegraphics[width=\textwidth]{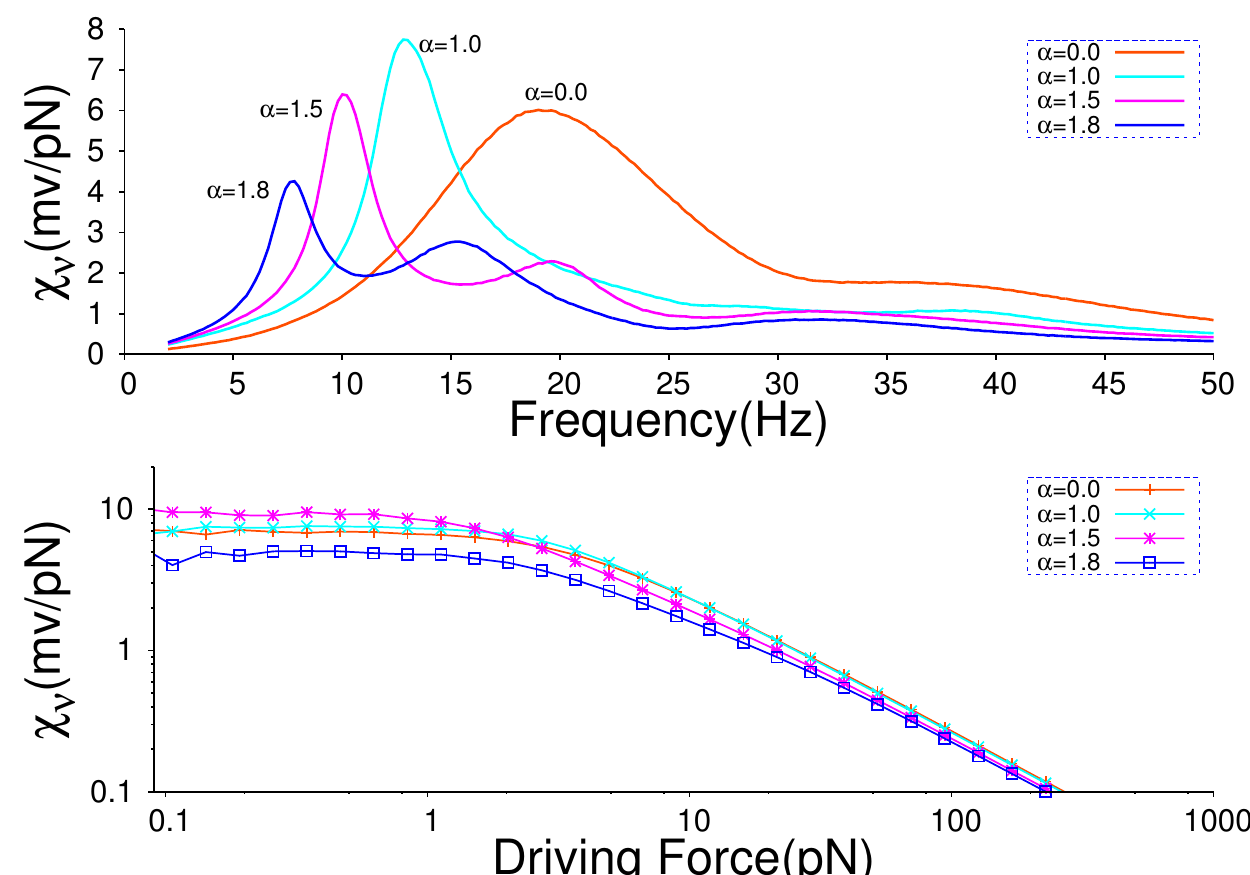}
\end{subfigure}
\caption{Effect of the backward coupling on sensitivity of the hair cell model to external periodic force. 
Left column shows the mechanical sensitivity, i.e. the hair bundle sensitivity, $\chi_f$. Right column shows the electrical sensitivity, $\chi_V$. Top panels shows sensitivity versus stimulus frequency, $f_\mathrm{s}$, for the fixed value of the stimulus amplitude $F_0=2$~pN and for indicated values of the backward coupling strength $\alpha$.  
Bottom panels show the sensitivity versus stimulus amplitude, $F_0$. For indicated value of $\alpha$ the stimulus frequency was chosen to match the frequency of spontaneous oscillation. Direct mechano-electrical coupling strength was fixed at $g_\mathrm{MET}=1$~nS.} 
\label{sens-lines}
\end{figure}

\noindent
Frequency response of mechanical and electrical compartments is shown in Fig.~\ref{sens-lines} for a weak mechanical sinusoidal stimulus (\ref{Fext}). Sensitivity of mechanical and electrical compartments peaks at the frequency of spontaneous oscillations. Furthermore, the peak values of sensitivity is largest for the coupling strength corresponding to the most coherent spontaneous oscillations (cf Fig.~\ref{op}, left panel). 
The sensitivity as a function of the stimulus strength shown in the lower panels of Fig.~\ref{sens-lines}, demonstrates the phenomenon of compressive nonlinearity. Both compartments respond linearly to weak stimuli, while the sensitivity to stronger stimuli is suppressed. 
The differences in response properties appear for strong stimuli $F_0 >100$~pN where the hair bundle start to respond linearly again, while the sensitivity of the electrical compartment continues to decline. We note that similar effect was reported in \cite{Neiman2011c} for voltage response where a linear model for the hair bundle was used. We also note that the backward coupling strength does not affect significantly the scaling of sensitivity with the stimulus amplitude. 
\begin{figure}
\centering
\begin{subfigure}[b]{0.45\textwidth}
  \includegraphics[width=\textwidth]{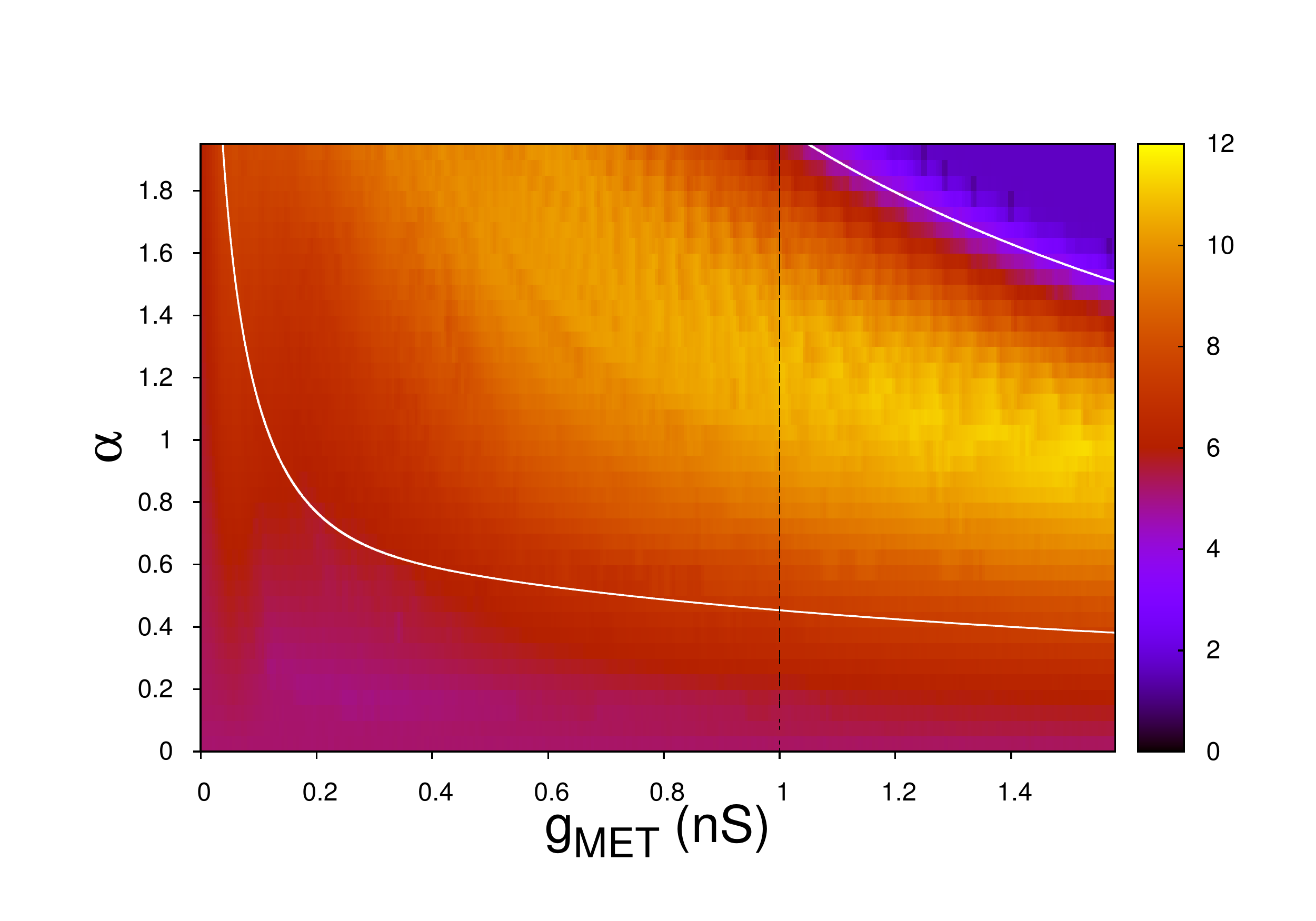}
\end{subfigure}
\begin{subfigure}[b]{0.45\textwidth}
 \includegraphics[width=\textwidth]{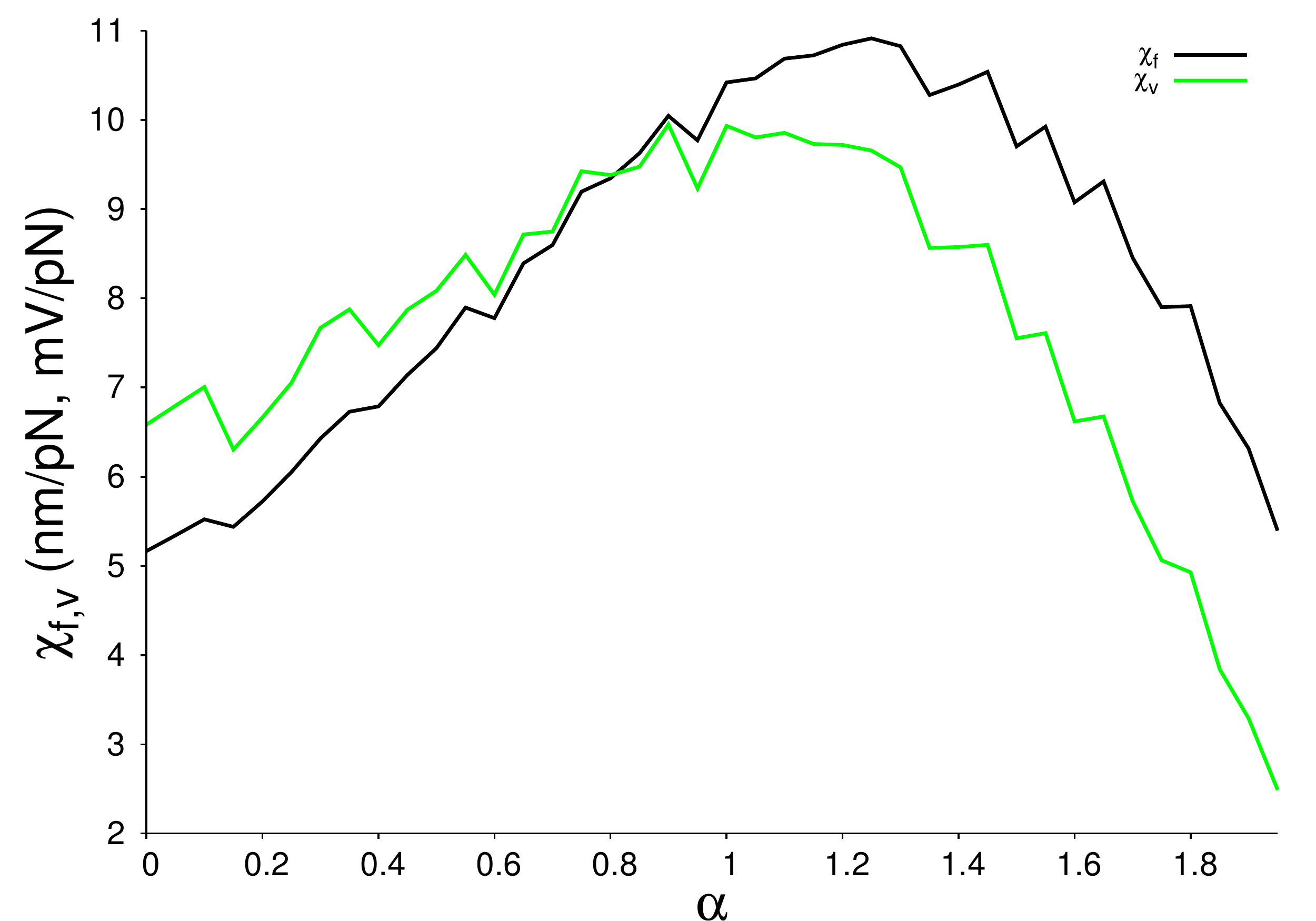}
\end{subfigure}
\caption{Maximal sensitivity of the hair cell model to weak sinusoidal force, $F_0=2$~pN. Left panel shows maximal sensitivity of the hair bundle versus both forward ($g_\mathrm{MET}$) and backward ($\alpha$) coupling strengths. Right panel shows sensitivities of the hair bundle and membrane potential versus $\alpha$ for the fixed value $g_\mathrm{MET}=1$~nS.}
\label{psmaps}
\end{figure}

The hair bundle sensitivity was studied for a wide range of forward and backward coupling strengths ($g_\mathrm{MET}\, \mathrm{\in [0,3]nS}$), ($\alpha \in [0,2]$). For each values of $g_\mathrm{MET}$ and $\alpha$ we estimated the maximum sensitivity by varying the stimulus frequency $f_\mathrm{s}$ and then plotted this maximum value of sensitivity as a color-coded map.  The resulting maximal sensitivity map is shown in Fig.~\ref{psmaps}, left panel. This map clearly shows the existence of optimal coupling strengths at which the hair cell is most sensitive to periodic perturbations.  For a fixed value of the  forward coupling strength $g_\mathrm{MET}=1.0$~nS   the sensitivity of the hair bundle peaks near $\alpha=1.5$, shown on the right panel of  Fig.~\ref{psmaps}, and corresponding well to the  most coherent spontaneous oscillation (see Fig.~\ref{op}).

\section*{Conclusions}
The role  of active processes in operational performance of sensory hair cells is a topic of intense current interest in sensory neuroscience \cite{Fettiplace2006a,Hudspeth2008,Fettiplace2009,Ashmore2010,Barral2011}. In particular, the role of active hair bundle dynamics is a matter of debate \cite{Hudspeth2010,Ashmore2010}. However, hair cells in lower vertebrates demonstrate also spontaneous oscillations of somatic potential which presumably may affect mechanics of the hair bundle, or may be a result of bi-directional coupling of mechanical and electrical constituents of the hair cell. Unlike many previous modeling works which studied mechanical and electrical dynamics separately,  we constructed a model  incorporating nonlinear mechanical and electrical compartments coupled bi-directionally. We showed that spontaneous oscillations may arise due to bi-directional coupling even when uncoupled compartments are quiescent. The coherence of oscillations can be enhanced by tuning the coupling strengths between compartments resulting in enhanced sensitivity and sharper tuning to weak periodic mechanical stimulus. At the same time, the model shows other distinctive behaviors, such as compressive nonlinearity.  

\bigskip
\noindent
{\bf Acknowledgments}\\
We thank B. Lindner and A. Shilnikov for fruitful discussions. This work was supported in part by the Quantitative Biology Institute at Ohio University.


\end{document}